\documentclass{ws-procs9x6-cpt16}

\begin{document}

\newcommand{\refeq}[1]{(\ref{#1})}
\def\etal {{\it et al.}}

\title{Proposed Test of Lorentz Invariance using the\\
Gravitational-Wave Interferometers
}

\author{A.C.\ Melissinos }

\address{Department of Physics and Astronomy, University of Rochester\\
Rochester, NY 14627, USA}

\begin{abstract}
Currently operating gravitational-wave interferometers are Michelson 
interferometers, with effective arm length $L\sim 4\times 10^5$ m. While the
interferometer remains in lock, data at the fsr sideband frequency encode
information on slow phase changes in the $f\sim 10^{-5}$ Hz
range, with a fringe sensitivity $\delta \phi /2 \pi \sim 10^{-10}$.
Preliminary LIGO data
presented in 2009  show no Lorentz violating
signal at the second harmonic of the Earth's sidereal rotation frequency.
This sets a limit on a possible change in refractive index, $\delta n/n < 2\times
10^{-22}$, an improvement of more than three orders of magnitude over existing
limits.
\end{abstract}

\phantom{}\vskip10pt\noindent
Present-day gravitational-wave detectors are Michelson interferometers
with the arms configured as Fabry-P\'erot cavities.
They are sensitive to gravitational waves with strains 
$h\sim 10^{-23}$ at frequencies $f \sim 100$ Hz.
They are also sensitive to changes in the effective refractive
index along the arms, even at frequencies in the $\mu$Hz range.

The interferometers are kept on a dark fringe and the observable signal,
which is the demodulated amplitude of the light reaching the dark port photodetector, is
directly proportional to the difference in the phase shift of the light
returning from the two arms,
\begin{equation} \Delta \phi = \delta \phi_1 - \delta \phi_2. \end{equation}
The fringe shift $\delta \phi_j$ in each arm,
\begin{equation} \frac{\delta \phi_j}{2 \pi} = \left( \frac{\delta L_j}{L}+ \frac{\delta f_c}{f_c} +\frac{\delta \bar n}{\bar n} \right) 
\frac{2L}{\lambda}, \end{equation}
depends on the arm length $L_j$, the frequency of the (carrier) light $f_c$,
and the effective refractive index, $\bar n$; $\lambda$ is the wavelength of
the carrier.

When the interferometer is ``locked" onto a dark fringe, $\Delta \phi =0$,
and this is achieved by adjusting both the (microscopic) arm length difference,
$\delta L = L_1 - L_2$ (modulo $\lambda /2$) as well as  the frequency of the carrier, $\delta f_c$;
clearly, $\delta \bar n$ is determined by external factors. Departures from $\Delta \phi = 0$
are recorded with high bandwidth (at 16.384 kHz) and this ``error
signal" is the output of the instrument. Similarly, the corrective 
actions of the servo mechanisms that return $\Delta \phi \rightarrow 0$ are
recorded and available for analysis.

Integration of the error signal over a sufficiently long time interval $T$, compared
to the servo response, yields 
\begin{equation} \int_{t-T/2}^{t+T/2} dt \frac{\Delta \phi}{2 \pi} \rightarrow 
\frac{2L}{\lambda} \int_{t-T/2}^{t+T/2} dt \frac{\Delta \bar n}{\bar n} \end{equation}
because when the interferometer is locked, the first two terms in Eq.\ (2) have been
returned to zero by the servo, while their fluctuations are stochastic and
average to zero. This shows the interferometers are sensitive to time-varying 
signals from Lorentz violation in the effective refractive index $\bar n$.

The Earth's sidereal rotation frequency is $f_{s}= 1.16058 \times 10^{-5}$ Hz,
while the annual rotation frequency $f_{a} = 3.16876 \times 10^{-8}$ Hz. Such 
frequencies are outside the band of interest for gravitational-wave searches, but
can be accessed by down-sampling the signals in the DARM-CTRL channel, which is an 
integral over the error signal. However it is preferable to consider a sideband channel 
displaced by the ``free spectral range" (fsr) frequency from the carrier, which,
serendipitously, circulates in the interferometer.

Note that the arms are in resonance when the carrier frequency $f_c$ takes the value 
$f_c = N f_{fsr}$, where $N$ is a large integer, of order $10^{10}$, and $f_{fsr}=
c/2L \approx 37.52$ kHz for LIGO. Resonance in the arms also occurs at the 
sidebands $f_{\pm} = f_c \pm f_{fsr}$, which when demodulated experience lower noise
than the demodulated carrier. Most importantly, given a {\it{macroscopic}} difference in arm 
lengths, $\Delta L = L_1 - L_2$, when the interferometer is in lock on the dark fringe,
the sideband $f_{+}$ is displaced from the dark fringe by a bias phase shift $\phi_B/2\pi =
\delta L/2 L \approx 3\times 10^{-6}$ per traversal of light. As a result, the power at 
the sideband frequency $f_{+}$, contains an {\it{interference term}} between the bias phase
shift and any phase shift resulting from changes in $\Delta \bar n$. This induces a 
modulation of the power at the fsr frequency seen in Fig.\ 2 of Ref.\ \refcite{MG12}.

After demodulation at the fsr frequency the power spectral density (PSD) is evaluated 
for 64-second-long data stretches, and is integrated in the range (37.52 $\pm$0.2) kHz.
The resulting time series for the S5 LIGO run, 
(March 31, 2006 to July 31, 2007) is shown in 
Fig.\ 2 of Ref.\ \refcite{MG12} where the twice yearly modulation is clearly evident; similar modulation, but
of lower depth, appears on a daily time scale and is indicated in the inset to the figure.

When the time series is spectrally analyzed, discrete lines appear at the known frequencies
of the daily and twice daily tidal lines.\cite{Melchior} The spectra are shown in 
Fig.\ 3  of Ref.\ \refcite{MG12} where the bandwidth resolution is $ BW = 2.4\times 10^{-8}$ Hz.
In the daily region, four lines are present and the observed frequencies match those
of the known $O_1, P_1,$ and $K_1$ lines. The dominant line at $f = 1.157\times 10^{-5}$ Hz
is at the daily {\it{solar}} frequency where no tidal component is present. Therefore it must
be attributed to human activity on a daily cycle.
In the twice daily region, four lines, $N_2, M_2, S_2$ and $K_2$ are expected, and are 
observed at their exact frequencies and with the observed power proportional to
the known tidal amplitude: this should be so since the tidal signal contributes an interference
term to the spectral power.

The spectral lines cannot arise from physical motion of the mirrors
because the interferometer is {\it{maintained in lock}}. However the tidal acceleration (force per unit mass)
has a horizontal component, typically $g_{\parallel} \sim 10^{-7} g \sim 10^{-6}$ m/s$^2$,
that varies in time at the tidal frequencies. Such a gravity gradient imposes a frequency
shift on the light propagating along the arm (and an inverse shift on the return trip). 
The resulting phase shift, for a single traversal, is \begin{equation} \delta \phi^{(j)}/2\pi=
(g_{\parallel}^{(j)}/\lambda) (L^2/c^2), \end{equation} where $(j)$ refers to the orientation
of the arm. We calculate $\delta \phi = \delta \phi^{(1)} - \delta \phi^{(2)}$ for the M2 
tidal line at the Hanford site, to find \begin{equation} \delta \phi^{(s)}/2\pi|_{M2} =
2\times 10^{-10}. \end{equation} Using a simulation of the interferometer\cite{Finesse}
the observed modulation of the data in the twice daily region reproduces qualitatively 
the above result. Thus we use Eq.\ (5) to calibrate all spectral lines.

Violation of Lorentz invariance revealed by the Earth's rotation would appear at the harmonics
of the sidereal frequency, primarily at the second harmonic. Furthermore, the data at
the second harmonic are free of human activity and all the tidal lines have exactly their 
predicted values, in both frequency and amplitude, lending confidence in the data.
The $n=2$ sidereal frequency coincides
with the $K2$ line  which has power $P(K_2) = 398$ counts,
while the tidal contribution is
$P_{K_2(\rm tidal)} = 438$ counts. The possible LV power is $ P_{LV} = (-40\pm 28)
{\rm{~counts}}$. Taking 2$\sigma$ as the upper limit, and using the $M2$ calibration we find for 
the phase shift and fractional change in $\bar n$ the result
\begin{equation}\left.\frac{\delta \phi^{(s)}}{2 \pi}\right|_{2 \omega} < 2\times 10^{-12}, 
\qquad
\left.\frac{\delta \bar n}{\bar n}\right|_{2\omega} = \left.\frac{\lambda}{2L}\frac{\delta \phi^{(s)}}{2 \pi}\right|_{2 \omega}
< 2.5 \times 10^{-22}.\end{equation} This result for $\delta \bar n /\bar n$ is an improvement by
more than three orders of magnitude over the most recently published value.\cite{Nagel}

The data in Fig.\ 2 of Ref.\ \refcite{MG12} show clear evidence for twice yearly modulation as quantitatively supported
by spectral analysis which yields for the observed modulation frequency  $ f_{\rm observed} =
(6.54\pm 0.6) \times 10^{-8}$ Hz, as compared to  $f_{2 \Omega} = 6.34\times 10^{-8}$ Hz, in agreement within 
the error.
The amplitude of the line implies a phase shift $ \delta \phi/2 \pi|_{2\Omega} = (3.2 \pm 0.2)\times 10^{-9} $
or equivalently\begin{equation} \delta \bar n/\bar n|_{2 \Omega} = ( 4\pm 0.25) \times 10^{-19}.\end{equation}
In the Standard-Model Extension\cite{Kostelecky}  this leads to
\cite{kmm}
$|\kappa_{tr}|= (3.1\pm 0.2)\times 10^{-9}$, whereas
the absence of a signal at $f_{\Omega}$ sets a limit $|\kappa_{tr}|< 0.9\times 10^{-9}$. The large value
of the twice annual modulation and the absence of a signal at $f_{\Omega}$ make it probable
that this ``anomalous" modulation is due to an instrumental effect. At this time
we can not find any obvious experimental causes for this anomaly, and the resolution of this issue 
will have to await new experimental results. For further details on this work see Ref.\ \refcite{kmm}.

\section*{Acknowledgments}
The preliminary data discussed here were obtained during the LIGO S5 run. I am indebted to the staff
and operators of the LIGO Hanford observatory and to the members of the LSC. In particular I thank W.E.\ Butler,
C.\ Forrest, T.\ Fricke, S.\ Giampanis, F.J.\ Raab, and D.\ Sigg who were closely involved with the design, implementation,
data taking and analysis of the fsr channel. I thank A.\ Kosteleck\'y and M.\ Mewes for providing the SME
coefficients for the LIGO configuration.

\end{document}